\documentclass{elsart}
\usepackage{graphicx}
\usepackage{bm} 

\journal{Physics Letters B}

\begin{document}

\begin{frontmatter}

\title{High resolution study of the $\Lambda p$ final state interaction in the reaction 
$p + p \to K^+ + (\Lambda p)$}
\author{The HIRES Collaboration:}
\author[c]{A.~Budzanowski},
\author[z]{A.~Chatterjee},
\author[o]{H.~Clement},
\author[o]{E.~Dorochkevitch},
\author[e]{P.~Hawranek},
\author[d] {F. Hinterberger}\corauth[cor]{Corresponding author}
\ead{fh@hiskp.uni-bonn.de},
\author[d]{R. Jahn},
\author[d]{R.~Joosten},
\author[f,w]{K.~Kilian},
\author[c]{S.~Kliczewski},
\author[f,w,m]{Da.~Kirillov},
\author[g]{Di.~Kirillov},
\author[h]{D.~Kolev},
\author[i]{M.~Kravcikova},
\author[e,f,w]{M.~Lesiak},
\author[f,w,m]{H.~Machner},
\author[e]{A.~Magiera},
\author[l]{G.~Martinska},
\author[g]{N.~Piskunov},
\author[f,w]{J.~Ritman},
\author[f,w]{P.~von Rossen},
\author[z]{B.~J.~Roy},
\author[d,w]{A.~Sibirtsev},
\author[g]{I.~Sitnik},
\author[c,d]{R.~Siudak},
\author[h]{R.~Tsenov},
\author[d]{K.~Ulbrich},
\author[l]{J.~Urban},
\author[o]{G. J.~Wagner}
\address[d]{Helmholtz-Institut f\"{u}r Strahlen- und Kernphysik, Universit\"{a}t Bonn, Bonn, Germany}
\address[g]{Laboratory for High Energies, JINR Dubna, Russia}
\address[m]{Fachbereich Physik, Universit\"{a}t Duisburg-Essen, Duisburg, Germany}
\address[f]{Institut f\"{u}r Kernphysik, Forschungszentrum J\"{u}lich, J\"{u}lich,
Germany}
\address[w]{J\"{u}lich Centre for Hadron Physics, Forschungszentrum J\"{u}lich, J\"{u}lich, Germany}
\address[i]{Technical University Kosice, Kosice, Slovakia}
\address[l]{P.J.~Safarik University, Kosice, Slovakia}
\address[c]{Institute of Nuclear Physics, Polish Academy of Sciences, Krakow, Poland}
\address[e]{Institute of Physics, Jagellonian University, Krakow, Poland}
\address[z]{Nuclear Physics Division, BARC, Mumbai, India}
\address[h]{Physics Faculty, University of Sofia, Sofia, Bulgaria}
\address[n]{University of Chemical Technology and Metallurgy, Sofia, Bulgaria}
\address[o]{Physikalisches Institut, Universit\"at T\"ubingen, Germany}

%
%
\begin{abstract}
The reaction $pp\to K^+ + (\Lambda p)$ 
was measured at $T_p=1.953$~GeV and $\Theta=0^{\circ}$ 
with a high missing mass resolution 
in order to study the $\Lambda p$ final state interaction. 
The large final state enhancement near the $\Lambda{p}$ threshold
can be described using the standard Jost-function approach.
The singlet and triplet scattering lengths and effective ranges are
deduced by fitting 
simultaneously the $\Lambda{p}$ invariant mass spectrum and the total
cross section data of the free $\Lambda{p}$ scattering.
\end{abstract}
\end{frontmatter}
%

\section{Introduction}
The study of the $\Lambda{p}$ interaction is 
part of the systematic investigation of 
the hyperon-nucleon ($YN$) interaction.
The $YN$ interaction is an ideal testing ground for
the role of strangeness in low and intermediate energy physics.
It is also
of great relevance in studies of the SU(3)$_{\rm flavor}$
symmetry.
In addition the $YN$ interaction is needed  for studies of hypernuclei.
Experimental information on the $\Lambda{p}$ interaction has been obtained from
$\Lambda{p}$ scattering experiments 
~\cite{fhint:ale68,fhint:sec68},
the binding energies of hypernuclei
~\cite{fhint:dal81,fhint:dov84}, 
and studies of the $\Lambda{p}$ final state interaction (FSI).
The $\Lambda p$ FSI was observed in strangeness transfer reactions
$K^- + d \to \pi^- + (\Lambda {p})$
~\cite{fhint:tai69}
and in reactions with associated strangeness production
$\pi^+ + d \to K^+ + (\Lambda {p})$ and $p + p \to K^+ + (\Lambda {p})$
~\cite{fhint:mel65,fhint:ree68,fhint:hog68,fhint:sie94,fhint:bal98,fhint:bil98,fhint:mag01,abd06}.
The first high resolution study of the reaction 
$p + p \to K^+ + (\Lambda {p})$
was performed at Saclay \cite{fhint:sie94}
with proton kinetic energies of 2.3 and 2.7~GeV and missing mass resolutions between 2 and 5 MeV
depending on the scattering angle. 
The FSI enhancement close to the $\Lambda p$ threshold was later  
analyzed \cite{hint04}   in terms of the inverse Jost function and the effective range approximation  
\cite{gold64}.
Recently a  new method based on dispersion theory was developed 
\cite{gasp04,gasp05} in order to extract
the scattering length and the effective range  and to estimate the theoretical error.

Theoretical studies of the $\Lambda {p}$ interaction with the
Nijmegen \cite{fhint:sto99} and
J\"ulich \cite{hai05} meson-exchange models predict
the potentials, phase shift parameters and
effective range
parameters. Recently, the  $\Lambda{p}$ interaction
has been studied using the chiral effective field theory 
\cite{pol06,hai09}.
Another topic of the $\Lambda {p}$ FSI is the prediction 
of a narrow S=-1 dibaryon resonance by the cloudy bag model
with an invariant mass of 2109~MeV \cite{aer85}.

\section{Experiment}

\begin{figure}[t!]
\begin{center}
\includegraphics[width=0.50\textwidth]{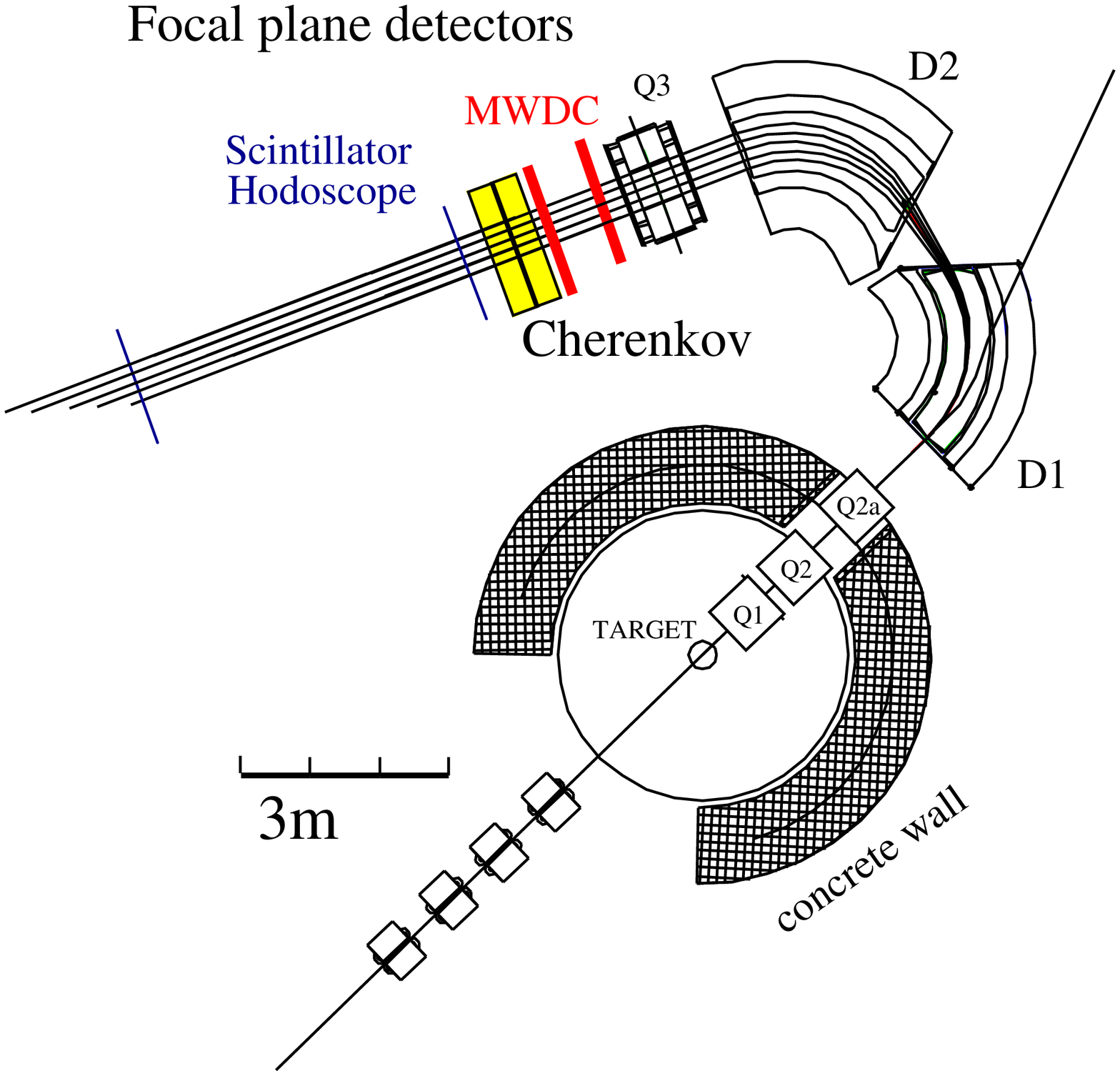}
\includegraphics[width=0.39\textwidth]{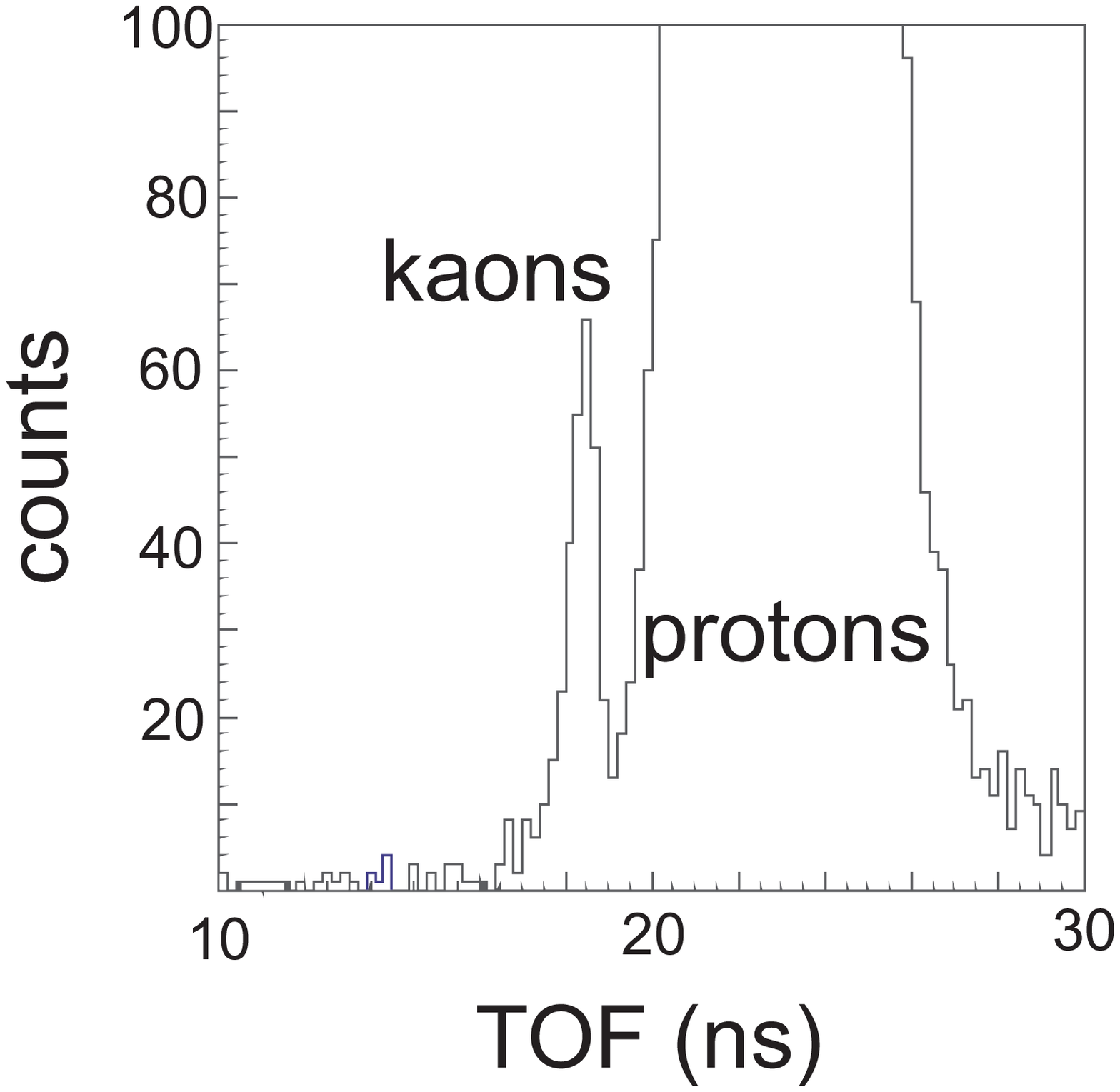}
\end{center}
\caption{Left: Layout of the magnetic spectrograph BIG KARL.
The charged particle tracks are measured in the focal
plane using  two stacks of multiwire drift chambers,
two threshold Cherenkov detectors and 
two scintillator hodoscopes. Right: TOF spectrum with pion suppression
by two Cherenkov detectors.}
\label{bigkarl}
\end{figure}

The reaction   $p + p \to K^+ + (\Lambda p)$
was measured at $0^{\circ}$
using the proton beam from the the cooler synchrotron COSY,
the  magnetic spectrograph BIG KARL \cite{bojo02}
and a 1.0~cm thick liquid hydrogen target (see Fig.~\ref{bigkarl}).
Since the most important part of the FSI enhancement is located in the
sharply rising part near the $\Lambda p$ threshold a
high missing-mass resolution was required.
The momentum of the incoming proton beam
was about 2.735~GeV/c corresponding to a kinetic energy of 1.953~GeV.
The absolute beam momentum was found from the fits of the kinematic parameters
for two reactions: $p + p \to K^+ + (\Lambda p)$ and $p + p \to d + \pi^+$
where kaons and deuterons were measured simultaneously at
BIG KARL momentum 1070~MeV/c.
The absolute precision of the beam
momentum was 0.15~MeV/c. 
The scattered  particles in the
momentum range  930 - 1110~MeV/c
were detected in the focal
plane using  two stacks of multiwire drift chambers,
two threshold Cherenkov detectors  and 
two scintillator hodoscopes.
The ratio of beam momentum to scattered particle momentum
was ideally suited for a measurement at $0^{\circ}$.
In the first dipole magnet of BIG KARL the beam was
magnetically separated from the scattered particles
and guided through the side exit of the outer yoke
into the beam dump. 
Thus, the huge background from dumping the
beam within the spectrometer has been avoided.
Particle identification was performed using the energy loss ($\Delta E$) and time of flight (TOF)
information from the scintillator hodoscopes.
In addition two threshold Cherenkov detectors \cite{siu08} were used in order to 
achieve a very high pion suppression factor of $10^5$.
The momentum of the kaon was measured and the missing mass of the $\Lambda p$ system was deduced. 
The effective missing mass resolution depending on the spread of the beam momentum, the momentum
resolution of the magnetic spectrograph and the 1~MeV bin width was $\sigma=0.84$~MeV.
In order to cover the missing mass range 2050 -- 2110~MeV three overlapping settings of the spectrograph 
(mean momenta: 1070, 1010 and 960~MeV/c) were used.
The relative precision of the momenta  of the three settings 
was 0.1 MeV/c.

\begin{figure}[h]
\begin{center}
\includegraphics[width=0.49\textwidth]{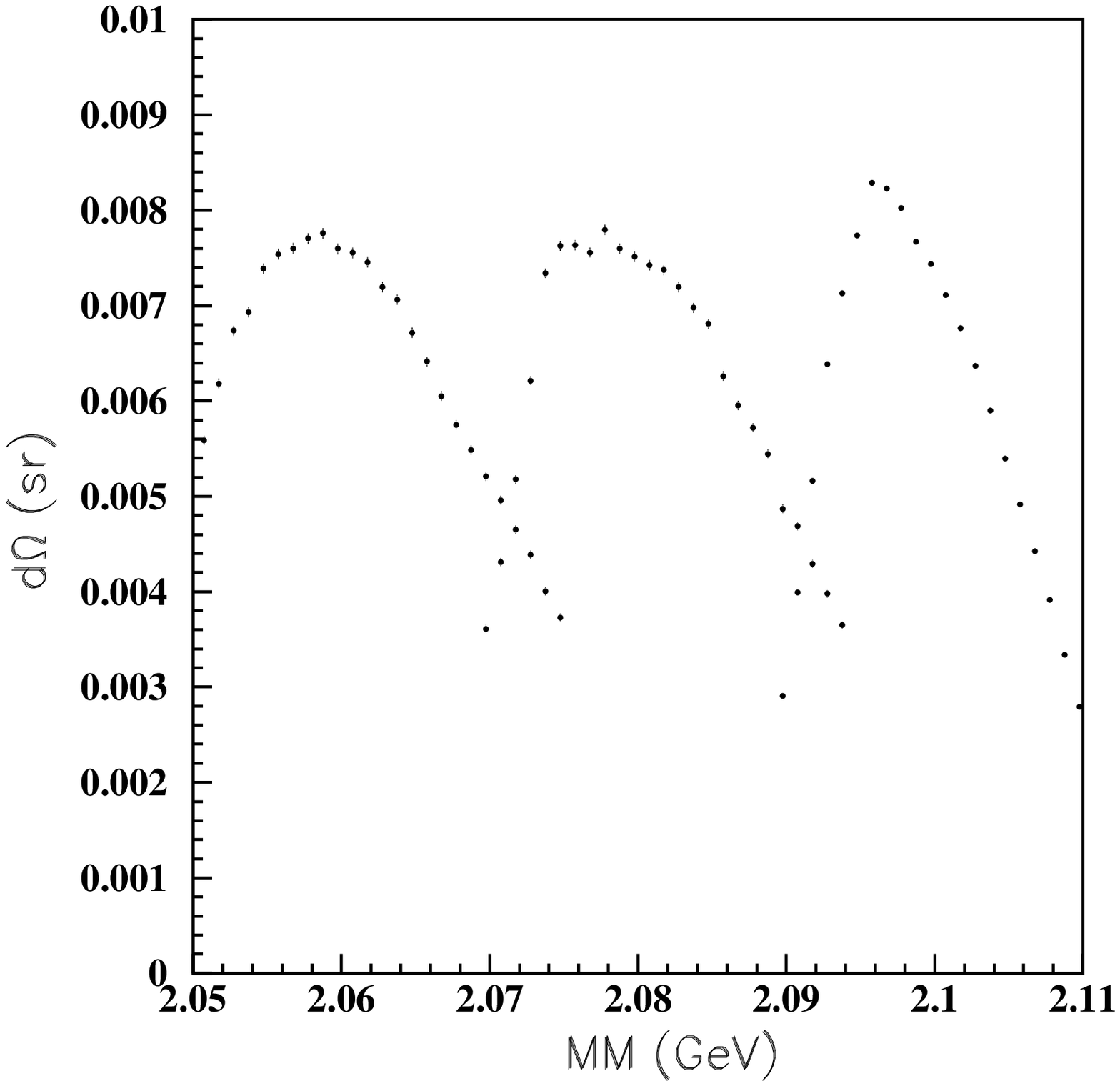}
\includegraphics[width=0.49\textwidth]{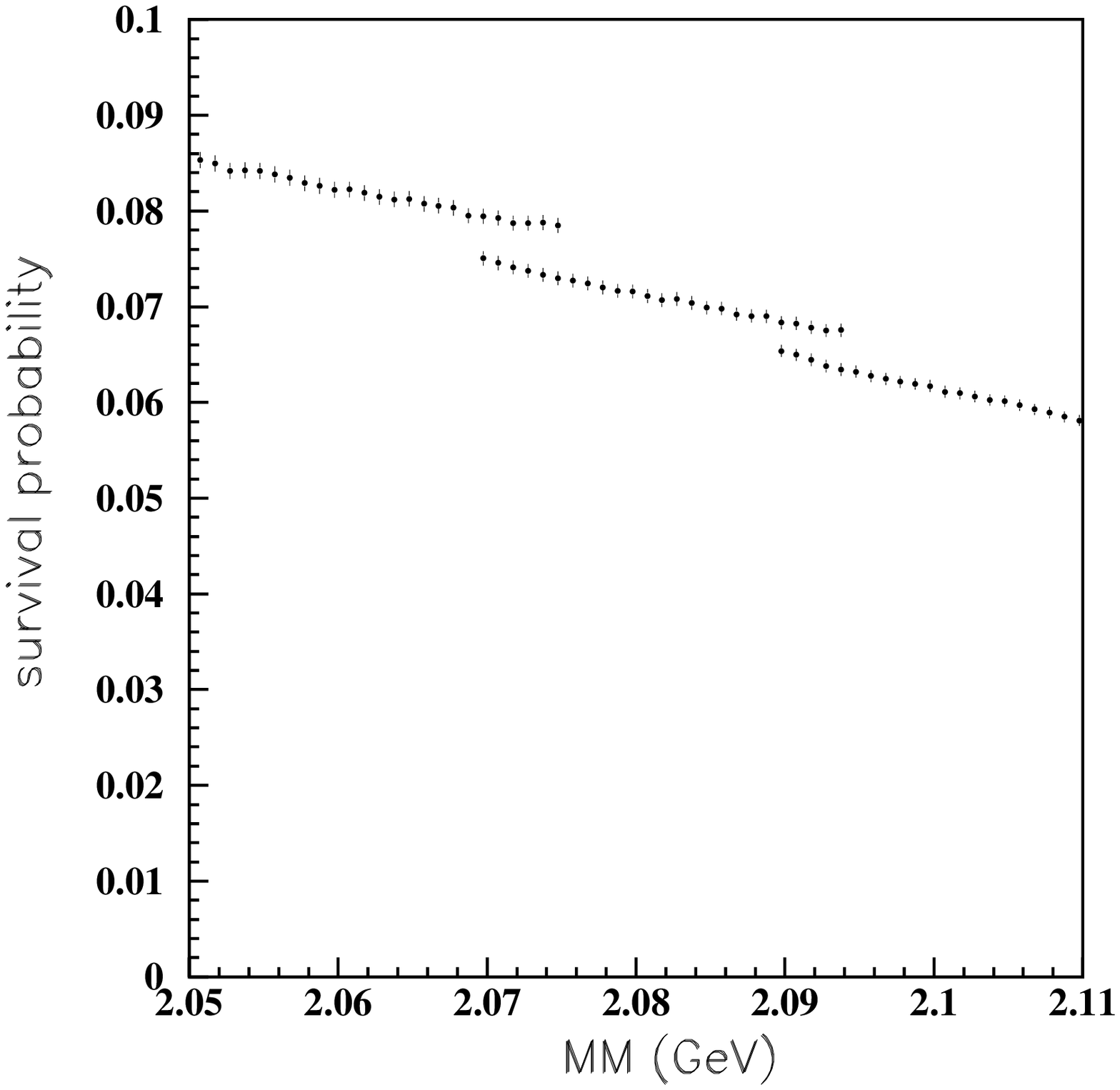}
\caption{Left: Acceptance correction functions. Right:
Kaon survival probability} 
\label{acceptance}
\end{center}
\end{figure}

Acceptance corrections with respect to solid angle and momentum were taken from
Monte Carlo calculations. 
The acceptance correction functions  contained  the magnetic spectrograph
momentum acceptance
around   $0^{\circ}$  with $(p_x,p_y,p_z)$ cuts
which for a given missing mass bin  
corresponded to the measured solid angle $d\Omega$. They contained also  
the detector efficiency corrections.
 The detector efficiency  included efficiency  of scintillator detectors which
 determined trigger and particle identification and
magnetic spectrograph efficiency with field inhomogeneity at the edges of the
acceptance.
Acceptance corrections as determined by Monte Carlo simulations were checked
by use of
experimental distributions of simultaneously 
measured  pions.
They are shown in Fig.~\ref{acceptance}.
The acceptance correction function of the 960~MeV/c setting looks
different due to a slightly different $(p_x,p_y,p_z)$ cut.

The kaon decay along flight path was taken into account for each individual
trajectory.
Path lengths obtained from  
calculations of tracks in the magnetic field
with the Turtle code \cite{Turtle}  were checked by experimentally deduced
values from time of
flight measurements of pions and protons.
The cross section error due to survival probability is less than 1~\%. 
The final kaon survival probability 
averaged for a given missing mass bin
is presented on the right side of Fig.~\ref{acceptance}.

The relative normalizations of the three different spectrograph settings were deduced from
luminosity monitors located in the target area
which were independent of the spectrograph settings.
The relative normalization errors 
due to the luminosity measurement were negligibly small. 
The relative normalization errors due to the acceptance
corrections  were estimated to be less than  2~\%.

\begin{figure}[h]
\begin{center}
\includegraphics[width=0.5\textwidth]{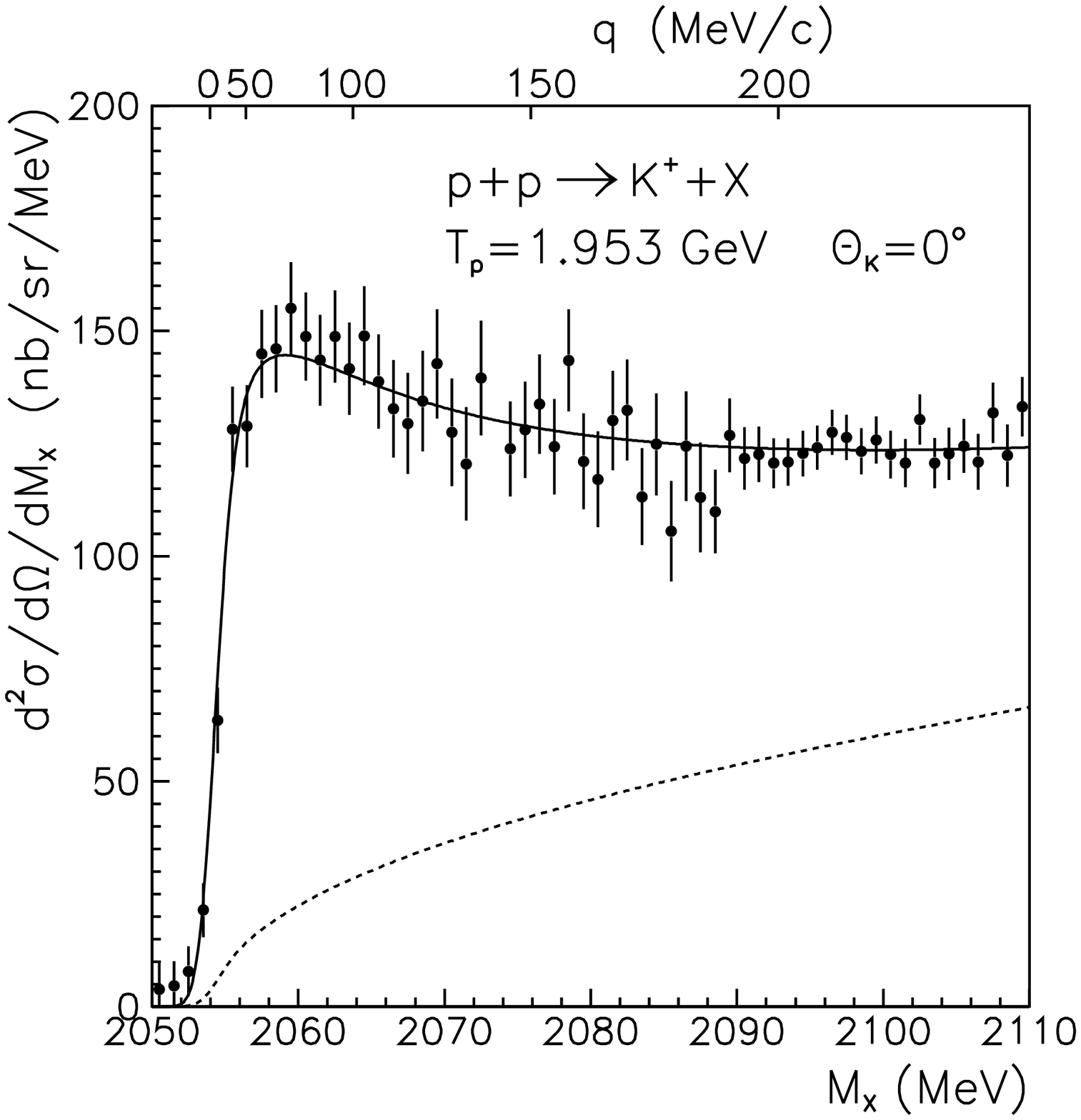}
\includegraphics[width=0.449\textwidth]{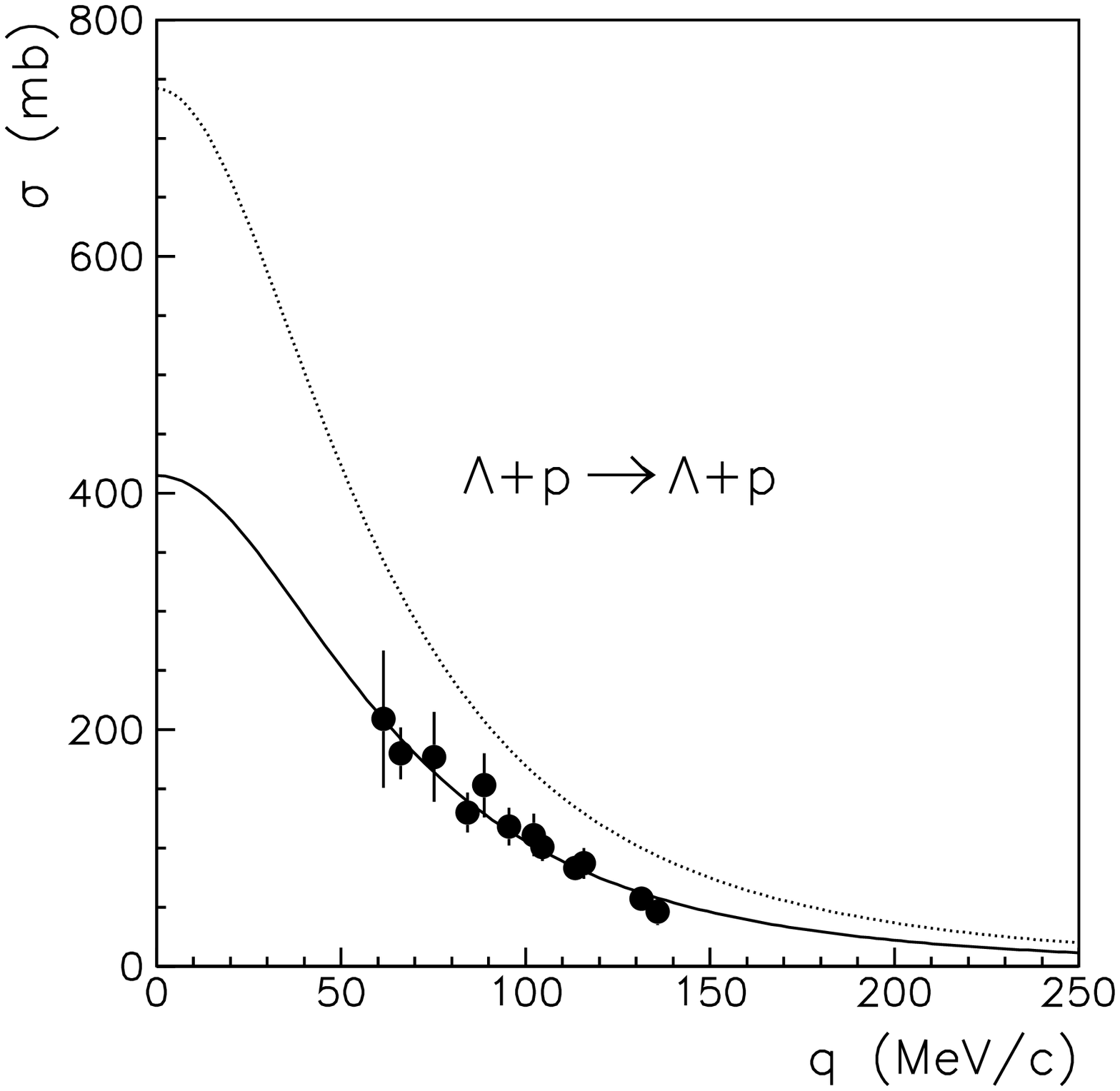}
\caption{Left: Missing mass spectrum of the reaction $p+p \to K^+ + (\Lambda p)$ 
measured at $T_p=1.953$~GeV and $\Theta_K=0^{\circ}$. 
The upper axis indicates 
the c.m. momentum $q$ of the $\Lambda p$ system.
Solid line: Combined six-parameter fit. 
Dashed line: $p+p \to K^+ + (\Lambda p)$ phase space distribution.
The region above 2090~MeV was measured with  
very small statistical errors in order to study a possible resonance anomaly
near 2096.5 and/or 2098.0 MeV \cite{fhint:sie94}.
Right: Total $\Lambda p \to \Lambda p$ cross section \cite{fhint:ale68,fhint:sec68} vs. c.m. momentum $q$.
Solid line: Combined six-parameter fit. Dotted line: Spin-averaged parameters.} 
\label{spectrum}
\end{center}
\end{figure}

The absolute cross section normalization was determined by measuring the 
luminosity  as described in \cite{bojo02}.
At the beginning of each beam period we calibrated two  luminosity monitors 
counting the left and right scattered particles 
from the target as function of the number of beam particles.
To this end, the beam current was highly reduced in order to
count the beam particles individually with a fast scintillator hodoscope in the beam dump.
For the calibration, the dependence of the luminosity signal on the beam intensity
was fitted by a linear function.
The resulting beam intensity error amounted to 5~\%. The density of the 
bubble-free liquid hydrogen target  ($\rho=0.0775$~g/cm$^3$)
with 1~$\mu$m thick  mylar foil windows was kept constant by stabilizing the
temperature to $15.0\pm 0.5$~K using
a high-precision temperature control  \cite{kilian02}.
The target thickness, i.e. the distance between entrance and exit window (nominal 1~cm) 
was precisely measured with a calibrated optical telescope.
The target thickness error was about 5~\%. 
The overall systematic normalization error was estimated to be  10~\%.
The missing mass spectrum is shown in Fig.~\ref{spectrum}.


\section{Formalism}

The observed missing mass spectrum can be described by factorizing the 
reaction amplitude in terms of a production amplitude
and a final state enhancement factor.
The method of parametrizing the FSI enhancement factor by the inverse Jost function \cite{gold64}
is described in \cite{hint04}. 
Taking the spin statistical weights into account 
the double differential cross section may be written as
\begin{equation}
\frac{d^2\sigma}{d\Omega_{K}dM_{\Lambda{p}}}= \Phi_3 \left[\, \frac{1}{4} \,
 |{M}_s|^2 \, \frac{q^2+\beta_s^2}{q^2+\alpha_s^2} 
+\, \frac{3}{4} \, |{M}_t|^2 \,
\frac{q^2+\beta_t^2}{q^2+\alpha_t^2}\right].
\label{miss}
\end{equation}
Here, $|{M}_s|^2$ and $|{M}_t|^2$ are the singlet and triplet production
matrix elements squared, $q$ the internal c.m.-momentum of the $\Lambda p$ system,
 $\alpha_s$, $\beta_s$, $\alpha_t$, $\beta_t$ the singlet and triplet potential parameters,
and $\Phi_3$ the ratio of the three-body phase space distribution\footnote{Note the 
difference of factor 2 in the definition of $\Phi_3$ with respect
to \cite{sib07}} and the incident flux factor.
The potential parameters $\alpha$ and $\beta$ can be used to establish
phase-equivalent Bargmann potentials~\cite{bar49b}.  
They are related
to the scattering lengths $a$, and effective ranges $r$ of the low-energy
$S$-wave scattering,
$\alpha=(1-\sqrt{1-2r/a})/r$,
$\beta=(1+\sqrt{1-2r/a})/r$.
Instead of the parameters
$\alpha_s$, $\beta_s$, $\alpha_t$ and $\beta_t$ one can equally well use the
singlet and triplet scattering length and effective range parameters $a_s$,
$r_s$, $a_t$ and $r_t$.  
The expression (\ref{miss}) is 
folded 
with the  Gaussian missing-mass
resolution function 
 ($\sigma = 0.84$~MeV)
before comparing with the data in the fit program.

The total
cross section $\sigma$ of the free $\Lambda{p}$ elastic scattering can be expressed
at low energies as a function of the c.m.-momentum $q$
using the effective range
approximation~\cite{gold64}, 
\begin{equation}
\sigma =\frac{1}{4}\, \frac{4\pi}{q^2{+}\left(-\frac{1}{a_s}{+}\frac{r_s
q^2}{2}\right)^2}
+ \frac{3}{4}\, \frac{4\pi}{q^2{+}\left(-\frac{1}{a_t}{+}\frac{r_t
q^2}{2}\right)^2}.
\label{cross}
\end{equation}

\section{Fit Results}

The nonlinear least-square fits are performed using the program Minuit
from the CERN library. 
We determine the spin-averaged scattering length $\bar{a}$
and effective range $\bar{r}$ 
in a three-parameter fit 
by applying the constraints
$|{M}_s|^2 = |{M}_t|^2{=}|\bar{{M}}|^2$,
$a_s {=} a_t {=} \bar{a}$, $r_s {=} r_t {=} \bar{r}$.
We fit only the missing mass spectrum
without taking the $\Lambda p$  total cross section data into account.
The fit yields an excellent description
of the missing mass spectrum with $\chi^2_{red}=0.55$,
$|\bar{M}|^2=27.8_{-2.0}^{+1.9}$~b/sr, $\bar{a}=-2.43_{-0.17}^{+0.16}$~fm and
$\bar{r}=2.21_{-0.16}^{+0.16}$~fm.
But these spin-averaged parameters
fail completely to reproduce the total $\Lambda p {\to} \Lambda p$
cross section data,
see dotted curve in Fig.~\ref{spectrum} on the right ($\chi^2_{red}{=}18$).
This result indicates that the 
singlet and triplet effective range parameters are different.

Therefore, we study the missing-mass spectrum and the total 
$\Lambda{p}$ cross sections simultaneously in a combined fit.
In addition we take a previous measurement of the $\Lambda p$ FSI in the reaction
$K^- + d \to \pi^- + p +\Lambda$ at rest into account \cite{fhint:tai69}
which yielded
an independent  determination of the triplet parameters, $a_t=-2.0\pm 0.5$~fm and
$r_t = 3.0\pm 1.0$~fm. These two values and their errors are taken as 
additional experimental data,
i.e. as 1-$\sigma$ constraints for the fitted $a_t$- and $r_t$-values
in the combined fit.
In order to explore the parameter space we start with a five-parameter fit
keeping the ratio $|M_t/M_s|^2$ fixed.
Starting with $|M_t/M_s|^2=1$,  we varied the ratio $|M_t/M_s|^2$ between 0 and 8.
The resulting $\chi^2$ values depend rather strongly on the 
ratio $|M_t/M_s|^2$.  
The optimum fit is achieved
with $|M_t/M_s|^2=0$. A six-parameter fit confirms this result. The final results are listed in Table~\ref{tab1}.
\begin{table*}[h!]
\begin{center}
\caption{Six-parameter fit results.}
\label{tab1}
\begin{tabular}{ccccccc}
\hline\noalign{\smallskip}
$|M_s|^2$ (b/sr) & $a_s$ (fm) & $r_s$ (fm) & 
$|M_t|^2$ (b/sr) & $a_t$ (fm) & $r_t$ (fm) &
$\chi^2_{red}$\\
\noalign{\smallskip}\hline\noalign{\smallskip}
$111_{-38}^{+8}$   & $ -2.43_{-0.25}^{+0.16}$ &  $2.21_{-0.36}^{+0.16}$ &
$0.0_{-0.0}^{+19}$ & $-1.56_{-0.22}^{+0.19}$ & $3.7_{-0.6}^{+0.6}$   & 0.53 \\
\noalign{\smallskip}\hline
\end{tabular}
\end{center}
\end{table*}

We note that a possible  theoretical uncertainty of the Jost-function approach may be  in the
order of 0.4~fm for the scattering length and even more for the
effective range as suggested by  the analysis of pseudodata \cite{gasp05}.
This aspect will be discussed in a further paper.

Concerning the experimental uncertainties, the overall
normalization error yields a systematic error of 10~\%
for the production matrix element ($|\bar{M}|^2$ and $|M_s|^2$).
The effective range parameters are not affected. They depend only on the 
relative shape of the
missing mass spectrum. 
The scattering length $a$ is mainly determined by the strongly rising part
of the spectrum near the $\Lambda p$ threshold. The effective range 
parameter $r$ depends mainly on the spectral distribution  towards higher
invariant masses.
The relative normalization errors between
the three parts of the spectrum 
yield a systematic error of 0.02~fm for the scattering length 
($\bar{a}$ and $a_s$) and
0.13~fm  for the effective range parameter ($\bar{r}$ and $r_s$).
These errors are smaller than the error estimates of the fit, but
they must be taken into account in the evaluation of the total error.
The precision of the beam momentum and the kaon momenta is so high
that the deduced FSI parameters are not affected.

The results shown in Table 1 depend on the accuracy of the
included $\Lambda p$ data, especially on the overall normalization
of the cross sections, see the right side of Fig.~\ref{spectrum}.
The error bars indicate the statistical errors which are the
main source of errors in those hydrogen bubble chamber experiments.
The systematic errors are small compared to the  statistical errors.
The data are based on 378 and 224 elastic $\Lambda p$
scattering events, respectively. They were 
taken by two independent groups \cite{fhint:ale68,fhint:sec68}
using
the 81-cm hydrogen bubble chamber at CERN. The two data sets are
consistent within the errors.

\section{Discussion and conclusion}

The reaction $p + p\to K^+ + (\Lambda p)$ 
was measured at $T_p=1.953$~GeV and $\Theta=0^{\circ}$ with a high missing mass resolution
in order to study the $\Lambda p$ FSI.
A three-parameter fit with spin-averaged effective range parameters
$\bar{a}$ and $\bar{r}$ yields a good description of the
missing mass spectrum but fails completely to describe
the momentum dependence of the free $\Lambda p$ scattering.
The combined study of the missing mass spectrum and the free  
$\Lambda p$ scattering using five- and six-parameter fits
reveals that the production of the 
$\Lambda p$ system in the triplet state is 
rather small.
The best fit yields $|M_t|^2 = 0$ and the $1\sigma$ limit is reached for
$|M_t/M_s|^2=0.168$
corresponding to $|M_s|^2 = 73.6$~b/sr and  $|M_t|^2 = 12.4$~b/sr.
In the fit we take also the independent determination of the 
triplet parameters $a_t$ and $r_t$ by Tai Ho Tan \cite{fhint:tai69} 
into account. 

It is interesting to note that the spin-averaged parameters
$\bar{a}$ and $\bar{r}$  
deduced from a three-parameter fit of the missing mass spectrum are
identical with the singlet parameters $a_s$ and $r_s$  
deduced from a six-parameter fit of both the missing mass spectrum
and the  free $\Lambda p$ scattering.
Also, the fit indicates that
the reaction $p + p\to K^+ + (\Lambda p)$ is dominated by
the singlet contribution,
however, this result should be considered with
respect to the theoretical uncertainties 
of the Jost-function approach \cite{gasp05}.
A direct determination of the
singlet and triplet contributions requires polarization
measurements as proposed in \cite{gasp04}.

The result $|a_s| > |a_t|$ indicates that
the $\Lambda p$
interaction is more attractive in the singlet state than in the triplet state.
This finding is in accordance with the analysis
of the binding energies of light hypernuclei \cite{fhint:dov84}.
The present results  agree within errors
with recent predictions of the new J\"ulich 
meson-exchange model J04c \cite{hai05} yielding
$a_s = -2.66$~fm,
$r_s = 2.67$~fm,
$ a_t = -1.57$~fm and
$r_t = 3.08$~fm.
Recent calculations 
of the  NLO effective field theory yield similar
results for $a_s$, $r_s$, $a_t$ and $r_t$
\cite{hai09}. 
The present results are also in agreement
with a similar analysis of the  reaction $p + p\to K^+ + (\Lambda p)$
at $T_p = 2300$~MeV and $\Theta = 10.3^{\circ}$ \cite{hint04}. 
Thus, the method of factorizing the reaction amplitude 
in terms of a production- and FSI-amplitude yields
comparable results for  
different bombarding energies and scattering angles.
In this sense, the present study provides also a systematic check
of the method.

We  analyze our data using the standard Jost-function approach \cite{gold64}.
The recently proposed dispersion-integral method \cite{gasp04,gasp05}
differs from the Jost-function approach 
by restricting the dispersion integral over the scattering phase shift 
to a finite upper limit.
In a further investigation we will 
analyze our data using different approaches.

A narrow  S=-1 resonance 
predicted by the cloudy bag model \cite{aer85}
is not visible in our missing mass spectrum.
The small structures at $2096.5\pm 1.5$ and $2098.0 \pm 1.5$~MeV
observed in a previous experiment \cite{fhint:sie94} are not confirmed. 
Further data analysis is underway in order to quantify
the resonance limits. But this is beyond the scope of the present paper.

%
%
\section*{Acknowledgements}\label{sec:Acknowl}
We thank 
the COSY team for excellent beam 
preparation.
This work was supported by the Bundesministerium f\"ur
Bildung und Forschung, BMBF (06BN108I), the Forschungszentrum
J\"ulich GmbH, COSY FFE (41520742), the European infrastructure
activity under the FP6 'Structuring the European Research
Area' programme, contract no. RII3-CT-2004-506078,
the Indo-German bilateral agreement and the GAS Slovakia (1/4010/07).

\end{document}